\documentclass[11pt]{amsart}
\usepackage{pgf, xcolor, url, amsmath, amsthm,sgame, tikz, color}
\usetikzlibrary{arrows,automata}
\usepackage[colorlinks=true, bookmarks=true]{hyperref}
   \hypersetup{colorlinks, citecolor=blue, filecolor=blue, linkcolor=blue, urlcolor=blue}
\usepackage[utf8]{inputenc}

\begin{document}
\title{Forecasting in light of Big Data}
\author{
Hykel Hosni         and
        Angelo Vulpiani
}

\address{ (H.H.) Dipartimento di Filosofia, Universit\`a degli Studi
  di Milano\\
and\\
(A.V.) Dipartimento di Fisica, Universit\`a degli Studi di Roma
  Sapienza and\
Centro Linceo  Inderdisciplinare ``Beniamino Segre'', Accademia dei Lincei, Roma
  (Italy).}
\email{hykel.hosni@unimi.it; Angelo.Vulpiani@roma1.infn.it}

\date{30 May 2017}

\begin{abstract}

Predicting the future state of a system has always 
been a natural motivation for science and practical applications.
Such a topic, beyond its obvious technical and societal
relevance, is also interesting from a conceptual point of
view. This owes to the fact that forecasting lends itself to two
equally radical, yet opposite methodologies. A reductionist
one, based  on the first principles,  and the na\"ive-inductivist one, based only on data. This latter view has recently gained
  some attention in response to the availability of unprecedented amounts of
  data and increasingly sophisticated algorithmic analytic techniques. The
  purpose of this note is to assess critically the role of \emph{big
    data} in reshaping the key aspects of forecasting and in particular the claim that \emph{bigger} data leads to
  \emph{better} predictions. Drawing on the  representative example of
  weather forecasts we argue that this is not generally the case.  We
  conclude by suggesting that a clever and context-dependent
  compromise between modelling and quantitative analysis stands out as
  the best forecasting strategy, as anticipated nearly a
  century ago by Richardson and von Neumann.


\end{abstract}

\maketitle

\begin{flushright} {\it Nothing is more practical than a good theory}
  (L. Boltzmann)
\end{flushright}

\section{Introduction and motivation}
\label{sec:intro}

Uncertainty spans our lives and forecasting is how we cope with it,
individually, socially, institutionally, and scientifically. As a
consequence, the concept of forecast is an articulate one.
Science, as a whole, moves forward by making and testing
forecasts. Political institutions make substantial use of economic
forecasting to devise their policies.  Most of us rely on weather
forecasts to plan our daily activities.  
Thus, in forecasting, the boundaries between the natural
and the social sciences are often crossed, as well as the boundaries
between the scientific, technological and ethical domains.

This rather complex picture has been enriched significantly, over the
past few years, by the rapidly increasing availability of methods for
collecting and processing  vast amounts
of data.  This revived a substantial interest in purely inductive
methods which are expected to serve the most disparate needs, from
commercial service to data-driven science. 
 Data brokers sell to third parties the digital footprints recorded by
 our internet
activities or credit card transactions.  Those can be put to a number
of different uses, not all of them ethically neutral. For instance, aggressive forms of personalised
marketing algorithms identify women who are likely to be pregnant
based on their internet activity, and similarly health related web
searches have been proved to influence individual credit scorings
\cite{Pasquale2015}. However, data-intensive projects lie at the heart
of extremely ambitious, cutting-edge scientific enterprises, including
the US ``Brain Research through Innovative Neurotechnologies''
(\url{http://www.braininitiative.nih.gov/}) and the African-Australian
Square Kilometer Array, a radio telescope array consisting of
thousands receivers (\url{http://skatelescope.org/}).\footnote{See,
  e.g. 
\cite{Casacuberta2014} for an appraisal of how, experiments of this
kind, may lead to a paradigm shift in the philosophy of science.}

Those examples illustrate clearly that \emph{big data} spans radically
diverse domains. This, together with its sodality with machine
learning, has recently been fuelling an all-encompassing enthusiasm, which is
loosely rooted on a twofold presupposition. First, the idea that \emph{big data}  will
 lead to \emph{much better} forecasts. Second, it will do so across the
board, from scientific discovery to medical, financial, commercial and political
applications. It is this enthusiasm which has recently led to making a case for the predictive-analytics
analogue of universal Turing machines, unblushingly referred to as The Master Algorithm \cite{Domingos2015}.

Based on this twofold presupposition, 
\emph{big data} and predictive analytics are
expected to have a major impact in society, in technology, and  all the way  up
to the scientific method itself
\cite{Mayer-Schonberger2013}. The extent to which those promises are
likely to be fulfilled is currently a matter for debate across a
number of disciplines \cite{Calude2016,Kitchin2014,Nowotny2016,Leonelli2016,Canali2016,Nural2015}, while some early success stories rather quickly turned into macroscopic failures \cite{Lazer2014}.
This note adds to the methodological debate by challenging both
  aspects of the presupposition for \emph{big data}
  enthusiasm. First, more data may lead to worse
    predictions. Second,  a suitably specified context is crucial for
    forecasts to be scientifically meaningful. Both
    points will be made with reference to a highly representative
    forecasting problem: weather predictions.

The remainder of the paper is organised as follows. Section ~\ref{sec:forecasting} begins by recalling that the
very meaning of scientific prediction depends significantly on an
underlying theoretical context. Then we move on, in 
Section \ref{sec:reductionist-view},  to challenging the  
  na\"ive inductivist view which goes hand in hand with \emph{big data}
  enthusiasm. In a rather elementary setting we illustrate the
  practical impossibility of inferring future behaviour from the past
  when the dimension of the problem is moderately large. Section \ref{sec:mother-all-forecasts}  develops this further by
emphasising that forecasts depend significantly on the modeller's ability to identify the proper
level of description of the target system. To this end we draw on the
history of weather forecasting, where the early attempts at arriving
at a quantitative solution turned out to be unsuccessful precisely
because they took into account \emph{too much data}. The
representativeness of the example suggests that this constitutes a serious challenge to the 
view according to which \emph{big data} could make do with the sole analysis
of correlations. 

The main lesson can be put as
follows:  as anticipated nearly a century ago by
Richardson and von Neumann, a clever and context-dependent trade-off between modelling and quantitative analysis stands out as the best strategy
for meaningful prediction.
This flies in the face of the by now infamous claim put forward in
2008 in \emph{Wired} by its then editor C. Anderson ``the data
deluge makes the scientific method obsolete''. In our experience
academics have a tendency to roll their eyes when confronted with
this, and similar claims, and hasten to add that non-academic
publications should not be given so much credit. We believe
otherwise. Indeed we think that the importance of the \emph{cultural}
consequences of such claims is reason enough for academics to take
scientific and methodological issue against them, independently of
their publication venue. Whilst Anderson's argument fails to stand
methodological scrutiny, as the present paper recalls, its key message
--big data enthusiasm-- has clearly percolated  society at
large. This may   lead to very serious social and ethical shortcomings. For the
combination of statistical methods and machine learning techniques for
predictive analytics is currently finding cavalier application in a number of very sensitive intelligence and policing
activities, as we now briefly recall. 

This clearly illustrates that
the scope of the epistemological problem tackled by this note extends far beyond the
scientific method and the academic silos.

\subsection{From SKYNET to PredPol}
\label{sec:from-skynet-predpol}
Early in 2016 a debate took place on alleged drone attacks in
Pakistan. The controversial article by C. Grothoff and
J.M. Porup\footnote{http://arstechnica.co.uk/security/2016/02/the-nsas-skynet-program-may-be-killing-thousands-of-innocent-people/}
opened as follows:

\begin{quote}
  In 2014, the former director of both the CIA and NSA proclaimed that
  ``we kill people based on metadata.'' Now, a new examination of
  previously published Snowden documents suggests that many of those
  people may have been innocent.
\end{quote}

Recall that SKYNET is the
US National Security Agency's programme aimed at monitoring mobile
phone networks in Pakistan. Leaked
documents \cite{Intercept} show that the primary goal of this programme is the identification of potential affiliates to the Al
Quaeda network. Further information suggests that SKYNET builds on classification techniques, fed
primarily 
on GSM data drawn from the entire Pakistani population. This obviously
puts the classification method at high risk of overfitting, given, of
course, that the vast majority of the population is not linked to
terrorist activities.  Not surprisingly then, the Snowden papers
revealed a rather telling result of the SKYNET sophisticated machine
learning, which led to attach Ahmad Zaidan, a bureau chief for
Al-Jazeera in Islamabad, the highest probability of being an Al Quaeda
courier. 

Two points are worth observing. First note, as some commentators have reported \cite{Robbins}, that 
the classification of Zaidan as strongly linked to Al Qaida cannot be
dismissed as utterly wrong. It of course all depends on what we
\emph{mean} by ``being linked''. As a journalist in the field he was
certainly ``linked'' to the organisation, and very much so if one
counts the two interviews he did with Osama Bin Laden. But of course,
``being linked'' with a terror organisation may mean something
entirely different, namely being actively involved in the pursuit of
their goal. This fundamental bit of contextual information is probably
impossible to infer for a classification technique, even the most accurate
one. But SKYNET algorithms are far from it,  which
brings us to the second noteworthy point.  The leaked documents assess the rate of false positives of the
classification method used by SKYNET between 0.008\% and 0.18\%. Since the surveillance
programme gathers data from a population of 55 million people, this
leads to up to 99 thousand Pakistani who may have been wrongly
labelled as ``terrorists''. Whether or not this actually led to deadly
attacks through the ``Find-Fix-Finish'' strategy based on Predator
drones, this example illustrates the shortcomings of the \emph{universality}
of the combination of big data and machine learning. For if the SKYNET
programme was about detecting unsolicited emails, rather than
potential terror suspects,  the false positive rate of 0.008\% would be considered exceptionally good. It is far from it, if it
may lead to causing highly defamatory accusations, if not outright
death to thousands of innocent people. The observation to the effect
that terrorists identification and spam detection are completely
different problems, with incomparable social, legal, and ethical
implications, though apparently trivial, may easily be overlooked as a
consequence of the big data enthusiasm.

On a less spectacular, but no less worrying scale, this can be seen to
feed the increasing excitement for \emph{predictive policing}. Police
departments in the United States and in Europe have been recently purchasing commercially available
software to predict crimes. California based  PredPol\footnote{ \url{http://www.predpol.com/}} is widely used across
  the country and by some police departments in the United
  Kingdom. The New York Times reports\footnote{\emph{The Risk to Civil
      Liberties of Fighting Crime With Big Data}, 6 November 2016}
  that Coplogic\footnote{\url{http://www.coplogic.com/}} has contracts
  with 5,000 police departments in the US. Keycrime\footnote{\url{http://www.keycrime.com/}} is a
  Milan based firm which has been recently contracted by the Italian
  police. This list can  be prolonged.  Predictive policing's main selling point is of course expense
reduction. If we can predict where the next crime is going
to be committed, we can optimise patrolling. Being more precise
requires less resources, less taxpayers money, and it delivers
surgical results. But context is once again neglected. When
introducing the methods and techniques underlying predictive policing
the authors of the 190 pages strong RAND report \cite{Perry2013} on the
subject note that 
\begin{quote}
  These analytical tools, and the IT that supports them, are largely
  developed by and for the commercial world.
\end{quote}

This, we believe, suffices to illustrate the relevance and
  urgency of a matter which we now move on to discuss in greater
  generality. To this end we shall begin by recalling a seemingly
  obvious, and yet surprisingly often 
overlooked, feature of the forecasting problem, namely that \emph{not all forecasts are equal}. 

\section{On forecasting}
\label{sec:forecasting}

Laplace grasped rather clearly one important feature of how probability
and uncertainty relate to information when he pointed out that
probability depends \emph{partly on our knowledge and partly on our
ignorance}. What we \emph{do} know clearly affects our understanding
of what we don't know and, consequently, our ability to estimate its probability.

It cannot be surprising then, that the meaning of scientific
prediction or forecasts changes with the growth of science.  In
\cite{Parisi1999}, 
for instance,  it is suggested that one can get a
clearer understanding of what physics \emph{is} by being specific about
the accepted meaning of physical predictions. 

The origins of the very concept of scientific forecast can in fact be
traced back to the beginning of modern physics.  The paradigmatic
example being classical mechanics -- the deterministic world in which  (for a limited
class of phenomena) one can submit definite Yes/No predictions to
experimental testing. A major
conceptual revolution took place in the mid 1800s with  the
introduction of
\emph{probabilistic prediction}, a notion which in the intervening two
centuries  has taken three distinct interpretations. The first relates to the
introduction of statistical mechanics, and is indeed responsible for
introducing a novel, stochastic, view of the laws of nature. The second started at the
beginning of the 1900s with the discovery of quantum mechanics. The
third, which is coming of age, relates to the investigation of complex
systems. It also observed in \cite{Parisi1999} that this development of the meaning of scientific forecasting amounted to its progressive weakening. Whilst the concept of stochastic prediction in
statistical mechanics is clearly \emph{weaker} than the Yes/No prediction of
the next solar eclipse, it can be regarded as being \emph{stronger}
than predictions about complex systems which may involve probability
intervals. The upside of increasingly weaker notions
of forecasts is the extension of the applicability of physics to a
wider set of problem. The downside is the lack of precision. 

It is interesting to note that the first major shift in
  perspective -- from the binary forecasts of classical mechanics to
  the probabilistic ones of statistical mechanics -- can be motivated
  from an informational point of view. To illustrate, we
  borrow from a classic  presentation of Ergodic Theory   \cite{Halmos}, in which a gas with $k$ molecules contained in a three-dimensional box is considered. Since particles can move in any
  direction of the (Euclidean) space, we are looking at a system with
  $n=3k$ degrees of freedom. Assuming complete information about the
  molecules' masses and the forces they exert, the
  \emph{instantaneous} state of the system can be described fully --at
  least in principle-- by
  fixing $n$ spatial coordinates and the $n$ corresponding velocities,
  i.e. by picking a point in  $2n-$dimensional Euclidean space. 
  We are now interested in looking at how the system evolves in time
  according to some underlying physical law. 
  \emph{In practice} though, the information we do possess is
  seldom enough to determine the answer.
  \begin{quote}
  [This led Gibbs to] abandon the deterministic
study of one state (i.e., one point in phase space) in favor of a statistical
study of an ensemble of states (i.e., a subset of phase space). Instead of asking ``what will the state of the system be at time $t$?'', we should ask
``what is the probability that at time $t$ the state of the system
will belong to a specified subset of phase space?''.\cite{Halmos}
  \end{quote}

This (to our present lights) very natural observation led to enormous
consequences. So it is likewise natural to ask, today, whether the
present ability to acquire, store, and analyse unprecedented amounts
of data may lead the concept of forecasts to the next level. 

In what follows we 
address this question in an elementary setting. In particular we ask
whether using our knowledge of the past states of a system  -- and without the
use of models for the evolution equation -- meaningful predictions
about the future are possible. Our answer is negative to the extent
that rather severe difficulties are immediately found, even in a very
abstract and simplified situation. As we shall point out the most 
difficult challenge to this view is understanding of the ``proper
level'' of abstraction of the system. This is apparent in the
paramount case of weather forecasting discussed in Section
\ref{sec:mother-all-forecasts}. We will see there that the key to
underestanding the ``proper level'' of abstraction lies with
identifying the ``relevant variables''  and the
effective equations which rule their time evolution.
It is important to stress that the procedure of  building such a 
description does not follow a fixed protocol, applicable in all
contexts given that certain conditions are met. It should rather be
considered as a sort of art, based on the intuition and the experience
of the researcher.

\section{An extreme inductivist approach to forecasting using Big Data}
\label{sec:reductionist-view}

According to a vaguely defined yet rather commonly held view \cite{Kitchin2014} \emph{big data}
may lead to dispense with theory, modelling or even hypothesising. All of this would be encompassed, across domains, by
smart enough machine learning algoritms operating on 
large enough data sets. This extreme inductivist
  conception of forecasts is thought of as depending solely on
  data. 
Is this providing us with a new meaning of predictions, and indeed one
which will outdate scientific modelling as we currently understand it? 

Two hypotheses, which are seldom made explicitly,  are needed to
articulate an  affirmative answer:
\begin{enumerate}
\item Similar premisses lead to similar conclusions (\emph{Analogy});
\item  Systems which exhibit a certain behaviour,  will continue doing
  so (\emph{Determinism}).
\end{enumerate}

Note that both assumptions are clearly at work in the
very idea of predictive policing recalled above. For predicting who is
going to commit the next crime and where this is going to happen, requires one to think
of the disposition to commit crimes as a persistent feature of certain
people, who in turn, tend to conform to certain specific
features. Those \emph{analogies} and the \emph{deterministic} character of the
`disposition to commit crimes' are very prone to mistake correlation
with causation. Racial profiling is the most obvious, but certainly
not the sole ethical concern which is being currently raised in
connection with the first performance assessments of predictive policing \cite{Saunders}.

Let us go back to our key point by noting that Analogy and Determinism
have been long debated in connection to
forecasting and scientific prediction. 

{\it If a system behaves in a certain way, it will do so again}
seems a rather natural claim, but, as pointed out by Maxwell\footnote{Quoted in Lewis Campbell and William Garnett, \emph{The Life of James Clerk
Maxwell}, Macmillan, London (1882); reprinted by Johnson Reprint, New York (1969), p. 440. } it is not
such an obvious assumption after all. 

\begin{quote} {\it It is a metaphysical doctrine that from the same
    antecedents follow the same consequents. [...] But it is not of much
    use in a world like this, in which the same antecedents never
    again concur, and nothing ever happens twice. [...] The physical
    axiom which has a somewhat similar aspect is ``That from like
    antecedents follow like consequents''.}
\end{quote}

In his \emph{Essai philosophique sur les probabilit\'es} Laplace  argued that analogy and
induction, along with a ``happy tact'', provide the principal means
for ``approaching certainty''  in situations in which  the
probabilities involved are ``impossible to submit to
calculus''. Laplace then  hastened to 
warn the reader against the subtleties of reasoning by induction  and the difficulties
of pinning down  the  right ``similarity'' between causes and effects which
is required for the sound application of analogical reasoning. 

More recently de Finetti sought to redo the foundations of probability
by challenging the very idea of  \emph{repeated} events, which
constitutes the starting point of frequentists approaches \emph{ a  la}
von Mises, a view  which is not central to Kolmogorov's
axiomatisation, but for which the Russian voiced some sympathy. In a
vein rather similar to that of Maxwell's, de Finetti argued
extensively \cite{DeFinetti1974,DeFinetti2008} that thinking of
events as ``repeatable'' is a modelling assumption.   If the modeller
thinks that two events are in fact instances of the same phenomenon, 
she/he should state \emph{that} as a subjective and explicit assumption.

This assumption is certainly not mentioned in the extreme inductivist  \emph{big data}
narrative, which advocates an approach to forecasting which uses just
knowledge of the past, without the aid of theory. 
Let us then turn our attention to this view, and frame
  the question in the simplest possible terms. We are interested in
  forecasts such that future states of a systems are predicted solely
  on the basis of known past states. If this turns out to be
  problematic in a highly abstract situation, then it can hardly be
  expected to work in contexts marred by high model-uncertainty, like
  the ones of interest for \emph{big data} applications.

Basically \cite{Time}, one looks for a past state
of the system ``near'' to the present one: if it can be found at
day $k$, then it makes sense to assume that tomorrow the
system will be ``near'' to day $k+1$. In more formal terms,
given the series $\{ {\bf x}_1, ... ,{\bf x}_M \}$ where 
${\bf x}_j$  is the vector describing the state of the system at time 
$j \Delta t$,  we look
in the past for an analogous state, that is a vector 
${\bf x}_k$ with $k <M$ ``near enough'' 
(i.e. such that  $|{\bf x}_k- {\bf x}_M|<\epsilon$, 
being $\epsilon$ the desired degree of accuracy). Once we
find such a vector, we ``predict'' the future at times
$M+n >M$  by simply assuming for ${\bf x}_{M+n}$  the state ${\bf x}_{k+n}$.
It all seems quite easy, but it is not at all obvious that an
analog can be found. 

The problem of finding an analog is strictly linked to
the celebrated Poincar\'e recurrence theorem\footnote{
In its original version the Poincar\'e recurrence theorem  states that:
\begin{quote} {\it Given a Hamiltonian system with a bounded phase
    space} $\Gamma$, {\it and a set} $ A \in \Gamma$, {\it all the
    trajectories starting from} ${\bf x} \in A$ {\it will return back
    to} $A$ {\it after some time repeatedly and infinitely many times,
    except for some of them in a set of zero probability.}
\end{quote}
Actually, though this is seldom stressed in elementary courses, 
the theorem can be  easily extended to dissipative ergodic systems provided one only considers 
initial conditions on the attractor, and ``zero probability" is interpreted with respect to the invariant probability 
on the attractor \cite{Collet}.
}: after a suitable time, a
deterministic system with a bounded phase space returns to
a state near to its initial condition \cite{Poincare,Kac}. Thus an analog  surely
exists, but how long do we have to go back to find it? The answer has
been given by the Polish mathematician Mark Kac who proved a Lemma  \cite{Kac}
to the effect that the average return time in a region $A$ is proportional to the inverse of the
probability $P(A)$ that the system is in $A$.

To understand how hard it is to observe a recurrence, and hence to
find an analog, consider a system of dimension $D$.\footnote{To be precise, if the
system is dissipative, $D$ is the fractal dimension $D_A$ of the
attractor \cite{Cecconi}.} The probability $P(A)$ of being in a region $A$ that
extends in every direction by a fraction $\epsilon$ is proportional to
$\epsilon^D$, therefore the mean recurrence time is $O(\epsilon^{-D})$ .
 If $D$ is large (say, larger
than $10$), even for not very high levels of precision (for
instance, $5\%$, that is $\epsilon=0.05$), the return time is so large
that \emph{in practice} a recurrence is never observed.

 That is to say that the required analog, whose existence
  is guaranteed in theory, sometimes cannot be expected to be found in
  practice, even if complete and precise information about the system
  is available to us.
  
\begin{figure}[t!]
\centering
\includegraphics[draft=false, scale=0.6, clip=true]{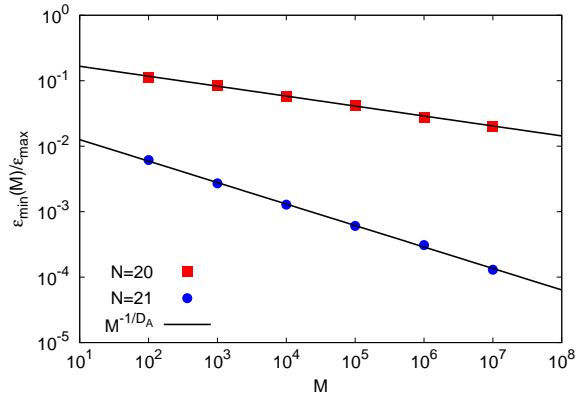}
\caption{
The relative precision of the best analog as function of the size $M$ of the
sequence.
The data have been numerically obtained\cite{Cecconi} from a simplified climatic model
introduced by Lorenz, with two different choices of the
parameters, see \cite{Lorenz}; the vector ${\bf x}$ is in $R^N$ with
$N=20$ and $N=21$.}
\label{Fig:1}
\end{figure}

Fig. \ref{Fig:1} shows how even for moderately large values of the
fractal dimension of the attractor $D_A$,   a good
analog can be obtained only in time series with enormous length.
If $D_A$ is small (in the example $D_A\simeq 3.1$) for an analog
with a precision $1 \%$ a sequence of length $O(10^2)$ is enough;
on the contrary for $D_A\simeq 6.6$ we need a very large sequence,
at least $O(10^9)$.

 In addition  usually we do not know the vector ${\bf x}$
describing the state of the system.
Such rather serious difficulty is well known in statistical physics; it has been
stressed e.g. by Onsager and Machlup \cite{Onsager} in their seminal
work on fluctuations and irreversible processes, with the caveat: 
{\it how do you
know you have taken enough variables, for it to be Markovian?};
and by  Ma \cite{Ma}: 
{\it the hidden worry of thermodynamics
is: we do not know how many coordinates or forces are necessary to
completely specify an equilibrium state.}

Takens \cite{Takens} gave an important contribution to such a topic:
he showed that from the study of a time series
$ \{ u_1, .. , u_M \}$,  where $u_j$ is an observable sampled at the
discrete times $j \Delta t$, it is possible (if we know
that the system is deterministic and is described by a finite
dimensional vector, and $M$ is large enough) to determine the proper
 variable ${\bf x}$.
Unfortunately, at practical level, the method has rather severe limitations: 

\begin{itemize}
\item[a)] It works only if we know \emph{a priori} that the system is
  deterministic;

\item[b)] The protocol fails if the dimension of the attractor is large
  enough (say more than $5$ or $6$).

\end{itemize}
Once again Kac's lemma sheds light on the key difficulty encountered here:
the minimum size of the time size $M$ allowing for the use
of Taken's approach increases as $C^M$ with $C=O(100)$ \cite{Time,Cecconi}.
 Therefore this  method cannot be used, apart for special cases
(with a small dimension), to build up a model from the
data. All extreme inductivist approaches will have to
  come to terms with this fundamental fact.
One of the few success of the method of the analogs is the 
tidal prediction from past history.
This in spite of the fact that tides are chaotic;
the reason is the low number of effective degrees of freedom
involved \cite{Cecconi}.

\section{Weather forecasting: the mother of all approaches to prediction}
\label{sec:mother-all-forecasts}
Weather forecasts provide a very good illustration of some central
aspects of predictive models. Not last because of the extreme accuracy
which this field managed to achieve over the past decades. And yet
this accuracy could be attained only when it became clear that
\emph{too much data} would be detrimental to the accuracy of the
model. Indeed, as we now briefly review, in the early days weather
forecasts featured a naive form of inductivism not dissimilar to the
one fuelling the big data enthusiasm. 

Let us stress that  the main limit to
predictions based on analogs is not the sensitivity to initial conditions, typical of chaos. But, as realized by Lorenz \cite{Cecconi},
the main issue is actually to find good  analogs.

The first modern steps in weather forecasting are due to 
Richardson \cite{Rich,Lynch} who, in his visionary work, 
introduced many of the ideas on
which modern meteorology is based. His approach was, to a certain
extent, in line with genuine reductionism, and may be summarised as
follows: the atmosphere evolves according to the hydrodynamic
(and thermodynamics) equations for the velocity, the density, and so on.
Therefore, future weather can be predicted, in principle at
least, by solving the proper partial differential equations, 
with initial conditions given by the present state of the atmosphere. 

The key idea by Richardson to forecast the weather was
correct, but in order to put it in practice it was necessary to
introduce one further ingredient that he could not possibly
have known \cite{Chibbaro}. 
After few decades von Neumann and Charney  noticed that the equations
originally proposed by Richardson, even though correct,
are not suitable for weather forecasting \cite{Lynch,DD}.
The apparently
paradoxical reason is that they are too accurate: they also
describe high-frequency wave motions that are irrelevant
for meteorology. So it is necessary to construct effective
equations that get rid of the fast variables. 

The effective equations have great  practical advantages,
e.g. it is possible to adopt large integration time steps  making the numerical computations satisfactorily efficient.
Even more importantly, they are able to capture the essence of the phenomena
of interest, which could otherwise be hidden in too detailed a
description, as in the case of the complete set of original
equations.
It is important to stress  that the effective equations
are not mere approximations of the original
 equations, and  they are obtained with a
subtle mixture of hypotheses, theory and observations \cite{DD,Chibbaro}.

\section{Concluding remarks}
\label{sec:concluding-remarks}
The above argument shows that in weather forecasting the accuracy of
prediction need not be monotonic
with the sheer amount of data. Indeed, beyond a certain point the opposite is
true. This, in our opinion, is a serious methodological objection to
the piecemeal \emph{big data} entusiasm. Given its representativeness among
all forecasting methods, the conclusions drawn with respect to
predicting the weather are far reaching, and help unifying a number of
observations that have been recently put forward along the same lines.

In many sciences and in engineering, an ever increasing gap  between
theory and experiment can be observed. This gap tends to widen
particularly in the presence of complex features in natural systems
science \cite{Parisi1999}. In socio-economical systems the gap between
data and our scientific ability to actually understanding them is
typically enormous. Surely the availability of  huge amounts of data,
sophisticated methods for its retrieval and unprecedented
computational power available for its analysis will undoubtedly help
moving science and technology forward. But in spite of a persistent emphasis on a \emph{fourth paradigm} (beyond the traditional ones,
i.e. experiment, theory and computation) based only on  data,
there is as yet no evidence data alone can bring about scientifically
meaningful advance. To the contrary, as nicely illustrated by
Crutchfield \cite{Crutchfield}, up to now  it seems that the unique way 
to understand  some  non trivial scientific or technological problem,
is following the traditional approach   based on a clever combination
of  data, theory (and/or computations),  intuition and wise use of
previous knowledge. Similar conclusions have been reached in the
computational 
biosciences. The authors of \cite{Coveney2016} point out very clearly
not only the methodological shortcomings (and ineffectiveness) of
relying on data alone, but also
unfold the implications of methodologically unwarranted \emph{big
  data} enthusiasm for the allocation of research funds to healthcare
related projects: ``A substantial portion of funding used to gather and process data should be diverted towards efforts to discern the laws of biology''.

\emph{Big data} undoubtedly constitute a great opportunity for scientific and
technological advance, with a potential for considerable
socio-economic impact. To make the most of it, however, the ensuing developments at the interface of
statistics, machine learning and artificial intelligence,
must be coupled with adequate methodological foundations. Not least because of the
serious  ethical, legal and more generally societal consequence of the
possible misuses of this technology. This note contributed to
elucidating the terms of this problem by focussing on the potential
for \emph{big data} to reshape our current understanding of
forecasting. To this end we pointed out, in a very elementary setting,
some serious problems that the na\"ive inductivist approach to
forecast must face: the idea according to which reliable
predictions can be obtained solely on the grounds of our knowledge of
the past faces insurmountable  
problems -- even in the most idealised and controlled modelling
setting.  

Chaos is often considered  the main limiting factor
to predictability in deterministic systems. 
However  this is  an unavoidable difficulty  as
long as the evolution laws of the system under consideration
are known. On the contrary, if the information on the system evolution 
is based only on observational data, the bottleneck lies
in Poincar\'e recurrences which, in turn, depend on the number of 
effective degrees of freedom involved. Indeed, even in
the most optimistic conditions, if the state vector of the system 
were known with arbitrary precision, the amount of data
necessary to make the meaningful predictions would grow
exponentially with the effective number of degrees of freedom, 
independently of the presence of chaos. However,
when, as for tidal predictions, the number of degrees of 
freedom associated with the scales of interest is relatively small,
the future can be successfully predicted from past history. In addition, in absence of a theory, a purely inductive
modelling methodology can only be  based on times series and the
method  on the analogs, with the already discussed difficulties
\cite{Time}. 

We therefore conclude that the \emph{big data} revolution is by all
means a welcome one for the new opportunities it opens. However the
role of modelling cannot be discounted: not only larger datasets, but
also the lack of an appropriate level of description \cite{DD,Chibbaro}
may make useful forecasting practically impossible.


\end{document}